\documentclass[aps,prb,twocolumn,groupedaddress]{revtex4}
\usepackage{graphicx}

\begin{document}
\title{ Pairing Symmetry of CeCoIn$_5$ Detected by In-plane Torque Measurements}
\author{H. Xiao, T. Hu, C. C. Almasan}
\affiliation{Department of Physics, Kent State University, Kent, Ohio, 44242, USA}
\author{T. A. Sayles, M. B. Maple}
\affiliation{Department of Physics, University of California at San Diego, La Jalla, California, 92903, USA}

\date{\today}
\begin{abstract}
In-plane torque measurements were performed on heavy fermion CeCoIn$_5$ single crystals in the temperature $T$ range $1.8$ K $\leq T \leq 10$ K and applied magnetic field $H$ up to 14 T. The normal-state torque is given by $\tau_n \propto H^4(1+T/T_K)^{-1}\sin 4\varphi$. The reversible part of the mixed-state torque, obtained after subtracting the corresponding normal state torque, shows also a four-fold symmetry. In addition, sharp peaks are present in the irreversible torque at angles of $\pi/$4, 3$\pi$/4, 5$\pi$/4, 7$\pi$/4, etc. Both the four-fold symmetry in the reversible torque and the sharp peaks in the irreversible torque  of the mixed state imply $d_{xy}$ symmetry of the superconducting order parameter. The field and temperature dependences of the reversible mixed-state  torque provide further evidence for $d_{xy}$ wave symmetry. The four-fold symmetry in the normal state has a different origin since it has different field and temperature dependences than the one in the mixed state. The possible reasons of the normal state four-fold symmetry are discussed.
\end{abstract}


\maketitle

\subsection{Introduction}
The superconductivity in the heavy fermion superconductor CeCoIn$_5$ is unconventional as exemplified by the non-Fermi liquid behavior,\cite{Petrovic, Sidorov, Kim} the giant Nernst effect present in the normal state, \cite{Bel} its proximity to quantum critical points, \cite{Ronning, Bianchi2} the Pauli limiting effect, \cite{Radovan, Bianchi1, Martin} and the possible multiband picture in the superconducting state.\cite{Rourke, Tanatar, Xiao} 
Unconventional superconductivity is always a subject of great interest. Knowing the pairing symmetry, which is related to the ground state and gap energy, is essential to the understanding of the pairing mechanism and the origin of superconductivity. For a  conventional superconductor, which is described by the BCS theory, the pairing is phonon mediated and the pairing symmetry is $s$ wave. For an unconventional superconductor, the quasiparticle gap vanishes at certain points on the Fermi surface. For example, most of the experimental evidence on high temperature superconductors, such as YBa$_2$Cu$_3$O$_{7-\delta}$, indicates that the $d_{x^{2}-y^{2}}$ wave symmetry dominates in these materials.

It has been established that the superconducting order parameter of CeCoIn$_5$ displays $d$-wave symmetry. $^{115}$In and $^{59}$Co nuclear magnetic resonance measurements\cite{Curro} and torque measurements\cite{Xiao2} have revealed a suppressed spin susceptibility, which implies singlet spin pairing.  A $T^{2}$ term is present in the low temperature $T$ specific heat, consistent with the presence of nodes in the superconducting energy gap.\cite{Movshovich} Nuclear quadrupole resonance and nuclear magnetic resonance measurements
on CeCoIn$_5$ have revealed that the nuclear spin lattice relaxation rate $1/T_1$ has no Hebel-Slichter
coherence peak just below the superconducting transition temperature $T_c$, and it has a   $T^{3}$ dependence at very low temperatures, which
indicates the existence of line nodes in the superconducting energy gap.\cite{Kohori}

Nevertheless, the direction of the gap nodes relative to the Brillouin zone axes, which determines the type of $d$-wave state, namely $d_{x^{2}-y^{2}}$ or $d_{xy}$, is still an open question and an extremely controversial issue. For example, angular dependent thermal conductivity measurements in a magnetic field have revealed four-fold symmetry consistent with $d_{x^{2}-y^{2}}$ symmetry.\cite{Izawa} Neutron scattering experiments by Eskildsen et al. revealed a square lattice oriented along [110], also consistent with $d_{x^{2}-y^{2}}$ wave symmetry. \cite{Eskildsen}  However, field-angle-dependent specific heat measurements  have found the symmetry of the superconducting gap to be $d_{xy}$.\cite{Aoki} Furthermore, theoretical calculations by Ikeda et al. strongly suggest $d_{xy}$ wave symmetry when taking into account the available $T^*$ experimental data of the FFLO states\cite{Ikeda2} [$T^*$ is the temperature at which the upper critical field $H_{c2}(T)$ changes from second order to first order]. In contrast, recent calculations by Tanaka et al. \cite{Tanaka} based on the Fermi liquid theory and by Vorontsov et al. \cite{Vorontsov} based on a unified microscopic approach,  support the $d_{x^{2}-y^{2}}$ gap symmetry. 

This extremely controversial issue needs to be resolved through an experimental technique that allows the direct measurement of the nodal positions. The experiments which are phase sensitive usually include surface or boundary effects, while the experiments which detect bulk properties are not phase sensitive. In the study presented here we use torque measurements to clarify the gap symmetry. Torque is a bulk measurement so it provides information on the order parameter of the bulk, not only the surface. (The order parameter of the surface might be different from that of the bulk). It also {\it directly} probes the nodal positions on the Fermi surface with high angular resolution since torque is the angular derivative of the free energy. Hence, such an experimental technique is ideal to determine the direction of the gap nodes relative to the Brillouin zone axes and, in fact, it has already been successfully used to identify the nodal positions of untwinned YBa$_2$Cu$_3$O$_{7-\delta}$ single crystals\cite{Ishida} and Tl$_2$Ba$_2$CuO$_{6+\delta}$ thin films.\cite{Willemin} In addition, recent theoretical calculations by H. Adachi have shown that low-field torque measurements can be used to detect the nodal positions of a $d$-wave superconductor with a small Fermi-surface anisotropy,\cite{Adachi3} as is the case of CeCoIn$_5$.  
 
In-plane torque measurements were performed on single crystals of CeCoIn$_5$ both in the normal state and in the mixed state. Normal-state torque shows a four-fold symmetry. The reversible part of the angular dependent mixed-state torque data, obtained after subtracting the corresponding normal state torque, also shows a four-fold symmetry with a positive  coefficient. The $T$ and $H$ dependences of the coefficients of the four-fold torques in the normal and mixed states are different. Hence, these two four-fold symmetries have clearly different origin. Sharp peaks in the irreversible torque data were observed at angles equal with $\pi/$4, 3$\pi$/4, 5$\pi$/4, and 7$\pi$/4, etc. The symmetry of the free energy extracted from the four-fold symmetry of the reversible torque of the mixed state coupled with the position of the sharp peaks in the irreversible torque of the mixed state point unambiguously towards  $d_{xy}$ wave symmetry in CeCoIn$_5$. Further support for the $d_{xy}$ wave symmetry is provided by the $H$ and $T$ dependence of the mixed-state reversible torque.

\subsection{Experimental Details}
Single crystals of CeCoIn$_5$ were grown using the flux method. High quality crystals with regular shape and shiny surfaces were chosen to carry out the torque measurements. The single crystals were etched in concentrated HCl for several hours to remove the indium left on the surface during the growth process. The crystals were then rinsed thoroughly in ethanol. 

A piezoresistive torque magnetometer was used to measure the angular dependence of the in-plane torque of CeCoIn$_5$ both in the normal state and mixed state. The torque was measured over a large temperature range (1.8 K $\leq T \leq$ 10 K)  and magnetic field $H$ range (1.5 T $\leq H \leq$ 14 T) by  rotating the single crystal in fixed magnetic field. The angle $\varphi$ for the in-plane rotation was defined as the angle made by the field with the $a$-axis of the single crystal.  
The contributions of the gravity and puck to the total torque signal were measured and subtracted from it as discussed elsewhere.\cite{Xiao}

The experiments were carried out in a Physical Property Measurement System (PPMS). In such a system with a one axis rotator, it is very difficult to ensure an in-plane alignment of better than about $\pm 3$ degrees. If misalignment exists, i.e. the magnetic field is not completely within the $ab$ plane of the single crystal, there  should be a $\sin 2\varphi$ term [see Eq. (6) of Ref. \cite{Aviani}]. Indeed, the angular-dependent in-plane torque signal has a $\sin 2\varphi$ term [$\alpha$ and $\varphi$ of Eq. (6) are $\varphi$ and $-23.7^0$, respectively, in the present case] in addition to the $\sin 4\varphi$ term. In fact,  the amplitude of the measured $\sin 2\varphi$ term gives a misalignment $\theta \approx 3.7^0$. So, we attribute this $\sin 2\varphi$ term to the misalignment of the single crystal.  The torque data shown in this paper are after subtracting this $\sin 2\varphi$ term. 
 
\subsection{Results and Discussion}
Previously, we have shown that the $b$-axis rotation torque signal measured in the mixed state has a paramagnetic component which is comparable with the diamagnetic component.\cite{Xiao} The former component is a result of the anisotropy of the susceptibilities along the $a$ and $c$ axes. Therefore, such a paramagnetic torque signal is absent in the present measurements  in which the torque is measured while rotating the single crystal along the $c$ axis, since $\chi_a \approx \chi_b$. Nevertheless, $T$ and $H$ dependent torque measurements in the normal state reveal that the normal-state torque signal is not negligible, is reversible, and it has a four-fould symmetry [see inset to Fig. 1(b)]. The solid line is a fit of the data with $\tau_{n}=A_n\sin 4\varphi$. The field and temperature dependence of the amplitude $A_n$, gives the $H$ and $T$ dependence of the normal-state torque. The coefficient  $A_n$ has an $H^{4}$ dependence up to 14 T for all measured temperatures from 1.9 K to 10 K [see Fig. 1(a)]. As the temperature increases, the slope of the plots in Fig. 1(a) decreases. This is consistent with the temperature dependence of $H^{4}/A_n$ shown in Fig. 1(b); i.e.,  $H^{4}/A_n$ increases, hence $A_n$ decreases, with increasing $T$. A straight line fit of the data with $H^{4}/A_n=c(1+T/T_0)$ gives $T_0=2.2$ K, which has a value close to the single ion Kondo temperature $T_K$ [reported to be between 1 and 2 K, Ref.\cite{Nakatsuji})] and $c=3\times 10^{10}$ T$^4$N$^{-1}$m$^{-1}$. Hence, $A_n(H,T) \propto H^4(1+T/T_K)^{-1}$. Therefore, in approaching the superconducting transition, $A_n$ decreases with decreasing $H$ and increases with decreasing $T$. We subtract the corresponding normal-state torque from the torque measured in the superconducting state.

The torque data measured in the mixed state of CeCoIn$_5$, obtained by subtracting the corresponding normal-state torque from the measured torque, have both reversible and irreversible components. The reversible torque $\tau_{rev}$ is the average of the torque data measured in clockwise and anti-clockwise directions, while the irreversible torque $\tau_{irr}$ is the average of the antisymmetric components of the torque data measured in clockwise and anti-clockwise directions.  
Figure 2(a) shows the reversible part of the angular dependent in-plane torque data measured in the mixed state at 1.9 K and in a magnetic field of 3 T. Clearly, there is a four-fold symmetry present in the torque data, although the data points are somewhat scattered. The solid line is a fit of these mixed-state data with $\tau_{rev}(H,T, \varphi)=A_{m}(T,H)\sin 4 \varphi$. The coefficient $A_m$ is positive since the torque displays a maximum at $\pi/8$. 
Figure 2(b) shows the irreversible part of the mixed-state torque data $\tau_{irr}(\varphi)$ measured at $T=1.9$ K and $H=1$ T for a single crystal with a mass of 2.6 mg. The $\tau_{irr}(\varphi)$ data have sharp peaks at $\pi/$4, 3$\pi$/4, 5$\pi$/4, and 7$\pi$/4, etc. 

The nodal positions of CeCoIn$_5$ can be obtained from the reversible and irreversible mixed-state torque data, as previously done in the study of YBa$_2$Cu$_3$O$_7$.\cite{Ishida} Specifically, theoretical calculations predict that the in-plane upper critical field $H_{c2}^{\parallel}$ has a four-fold symmetry for a $d$-wave superconductor.\cite{Takanaka}  In the case of $d_{xy}$ wave symmetry, the angular variation of the upper critical field $\Delta H_{c2}^{\parallel} \propto -\cos 4 \varphi$; hence, it has maxima at $\pi/$4, 3$\pi$/4, 5$\pi$/4, 7$\pi$/4, etc.  Figure 3 shows the angular dependence of the reversible and irreversible torque obtained by starting from this angular dependence of $H_{c2}$, as follows. The lower critical field $H_{c1}^{\parallel}$ is out of phase with $H_{c2}^{\parallel}$  (see Fig. 3) since the thermodynamic critical field $H^{2}_c = H_{c1}H_{c2} $ is independent of the magnetic  field orientation.  Therefore, the magnetization $M$, given by M$\simeq-H_{c1}ln(H_{c2}/H)/ln\kappa$,  has the same angular dependence as $H_{c2}$ (see Fig. 3).  
The easy axis of magnetization (maximum magnetization) should correspond to free energy $F$ minima. This implies that, for the $d_{xy}$ symmetry, $F$ has minima at $\pi/$4, 3$\pi$/4, 5$\pi$/4, 7$\pi$/4, etc. (see Fig. 3). The torque is the angular derivative of the free energy $F$; i.e.,  $\tau=-\partial F/\partial \varphi$. Hence, the reversible torque data for a material  with $d_{xy}$ wave symmetry should display a four-fold symmetry with maxima at $\pi/8$, $5\pi/8$, $9\pi/8$, etc (see Fig. 3). Also, the free energy minima act as intrinsic pinning centers for vortices, so the irreversible torque data for a material  with $d_{xy}$ wave symmetry should display peaks at the same angles at which the free energy has minima; i.e., at $\pi/4$, 3$\pi$/4, 5$\pi$/4, 7$\pi$/4, etc. Notice that the angular dependence of reversible and irreversible torque data of CeCoIn$_5$ shown in Figs. 2(a) and 2(b) is the same as the $\varphi$ dependence of $\tau_{rev}$ and $\tau_{irr}$, respectively, shown in Fig. 3, obtained from the theoretically predicted angular dependence of the upper critical fields for a material with $d_{xy}$ symmetry.  Therefore, the reversible along with the irreversible torque data in the mixed state unambiguously imply that the wave symmetry of CeCoIn$_5$ is $d_{xy}$.

To further understand the four-fold symmetry displayed by the present torque measurements in the mixed state, we studied the field and temperature dependence of the amplitude $A_m$.
Figure 4 is a plot of the $H$ dependence of $A_m$, which gives the $H$ dependence of the torque, obtained by fitting the angular dependent torque data measured in different magnetic fields. Note that $A_m(H)$ increases with increasing $H$, reaches a maximum, and then decreases with further increasing $H$.  Also note that the four-fold symmetry vanishes close to $H_{c2}^{\parallel}$ ($H_{c2}^{\parallel}=6$ T). This field dependence of the magnitude of $A_m$ is the same as the field dependence of the basal-plane reversible torque in the mixed state of a layered $d_{x^2-y^2}$ wave superconductor [see Fig. 8(a) of Ref. \cite{Adachi3}] calculated by Adachi et al.\cite{Adachi3}  based on the quasiclassical version of the BCS-Gor'kov theory with a Fermi surface which is isotropic  within the basal plane. The sign difference between the data of the present Fig. 4 and Fig. 8(a) of Ref. \cite{Adachi3}, which is for a $d_{x^2-y^2}$ wave symmetry, further indicates that the present data reflect $d_{xy}$ symmetry since the torque data have opposite signs for the $d_{xy}$ and $d_{x^2-y^2}$ wave symmetries.

The inset to Fig. 4 is a plot of the temperature dependence of the amplitude $A_m$. Note that $A_m$, hence the torque, decreases with increasing $T$ and vanishes towards $T_c$.  The fact that both the $T$ and $H$ dependences of the reversible mixed-state torque  vanish at the superconducting - normal state phase boundary further indicates that the observed four-fold symmetry is related with superconductivity; hence, it reflects the gap symmetry. 

We note that the behaviors of $A_m(H,T)$ and $A_n(H,T)$ are totally different (compare Figs. 1 and 4). So the four-fold symmetries present in normal and mixed states have different origin. The origin of the four-fold symmetry in the normal state is not yet clear to us. 

Torque measurements on LaCoIn$_5$ single crystals, which also have a tetragonal structure but are not superconducting and the $f$ electrons are absent, give some clues on the normal state four-fold symmetry of CeCoIn$_5$. The angular dependent torque data for LaCoIn$_5$ are shown in Fig. 5. Clearly, the $\sin 4\varphi$ symmetry observed in CeCoIn$_5$ is completely absent here. The difference in the normal-state torque between CeCoIn$_5$ and LaCoIn$_5$ could be due to the presence of heavy electrons in the former compound and their absence in the latter one. The crystalline electric field, which is important in heavy fermion systems, might be responsible for the four-fold symmetry in the normal-state torque. Also, the field induced order, possibly quadrupolar order, could be another reason for the four-fold symmetry in the normal-state torque of CeCoIn$_5$. No doubt, the origin of this normal-state four-fold symmetry present in the torque data requires further study. Nevertheless, this is beyond the scope of this paper.

\subsection{Summary}
In-plane angular dependent torque measurements were performed on CeCoIn$_5$ single crystals both in the normal and mixed states. Normal-state torque measurements show a four-fold symmetry. The reversible torque in the mixed state, obtained after subtracting the corresponding normal state contribution, also shows a four-fold symmetry with maxima at $\pi/8$, $5\pi/8$, $9\pi/8$, etc. Sharp peaks in the irreversible torque data were observed at $\pi/4$, $3\pi/4$, $5\pi/4$, etc. These latter peaks correspond to minima in the free energy of a $d_{xy}$ wave symmetry. The mixed state four-fold symmetry and the peak positions in the irreversible torque point unambiguously towards $d_{xy}$ wave symmetry of the superconducting gap.  The field and temperature dependences of the amplitude of the normal-state torque is different from that of the mixed-state torque, which indicates that the normal-state torque has a different origin. 
\\
\\
\textbf{Acknowledgments}
The authors would like to thank Professor K. Machida for fruitful discussions.
This research was supported by the National Science Foundation under Grant No. DMR-0406471 at KSU and the US Department of Energy under Grant No. DE-FG02-04ER46105 at UCSD. \label{} 
\\

\section{Figure Captions}
Figure 1. (Color online) Field $H$ and temperature $T$ dependence of the amplitude $A_n$ of the normal-state torque of CeCoIn$_5$ single crystals. (a) $H$ dependence of $A_n$ at measured temperatures of 1.9, 3, 4.5, 5, 6, 8, and 10 K. (b) $T$ dependence of $A_n$ at measured magnetic field of 14 T.  Inset: Angular $\varphi$ dependent torque $\tau_n$ measured in the normal state at $1.9$ K and  7 T in CeCoIn$_5$ single  crystals. The solid line is a fit of the data with $\tau_n=A_n \sin 4 \varphi$.

Figure 2. (a) Angular $\varphi$ dependence of the reversible  torque $\tau_{rev}$ measured at 1.9 K and  3 T on CeCoIn$_5$ single crystals. The solid line is a fit of the data with $\tau=A\sin 4 \varphi$. (b) Angular dependence of the irreversible torque $\tau_{irr}$ measured at 1.9 K and 1 T on CeCoIn$_5$ single crystals. Sharp peaks are present at $\pi/4$, $3\pi/4$, $5\pi/4$ and $7\pi/4$.

Figure 3. Plot of the angular $\varphi$ dependence of the upper critical field $H_{c2}$ and lower critical field $H_{c1}$ for magnetic field parallel to the $ab$ plane, magnetization $M$, free energy $F$, reversible torque $\tau_{rev}$, and irreversible torque $\tau_{irr}$ for $d_{xy}$  wave symmetry.

Figure 4. Field $H$ dependence of the amplitude $A_m$ of the reversible torque in the mixed state of CeCoIn$_5$ single crystals measured at $T=1.8$ K. The solid line is a guide to the eye. Inset: Temperature $T$ dependence of $A_m$ measured at $H=3$ T. The solid line is a guide to the eye.

Figure 5. Angular $\varphi$ dependent torque $\tau$ measured at 1.9 K and 14 T on LaCoIn$_5$ single crystals.

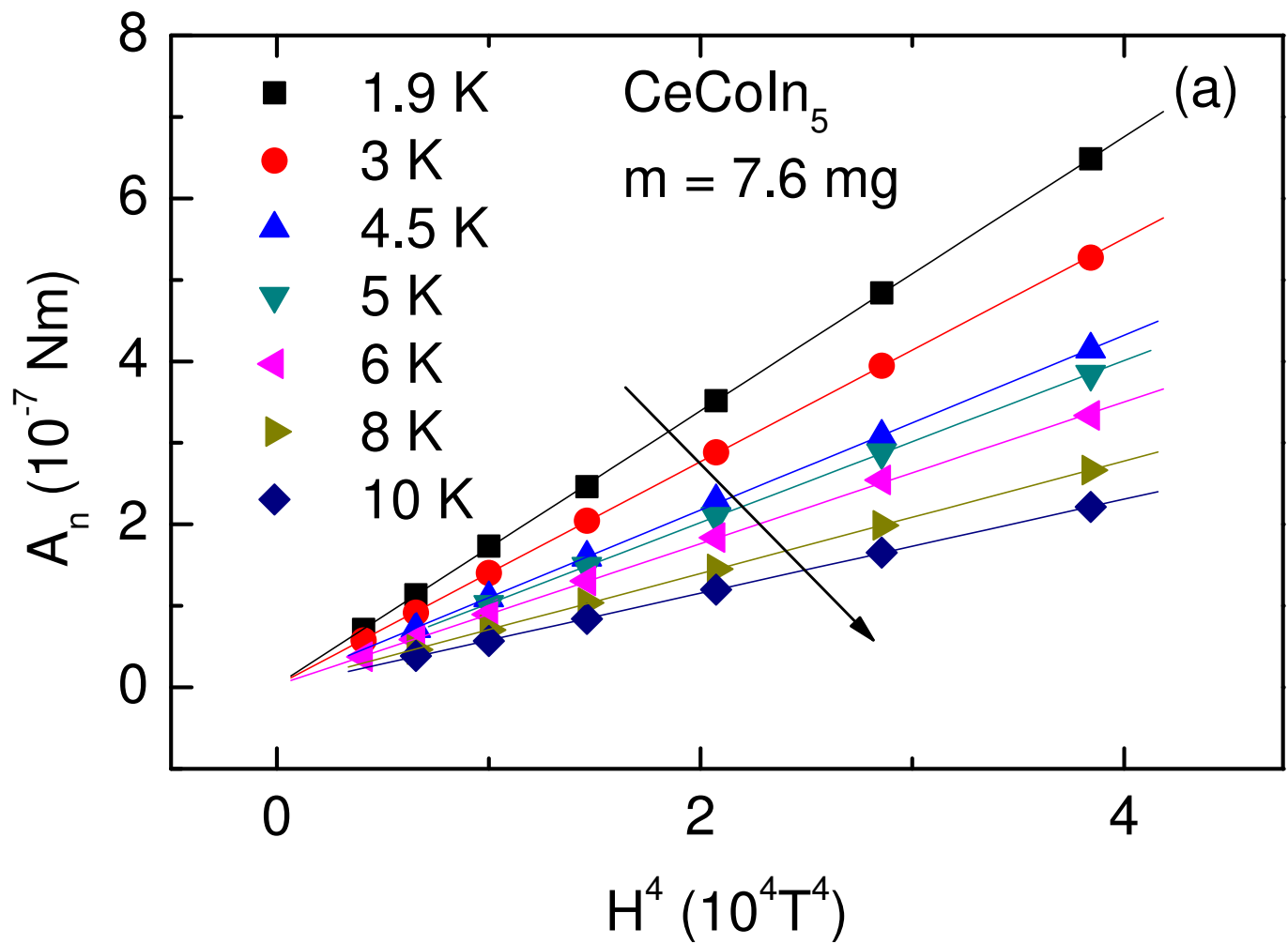
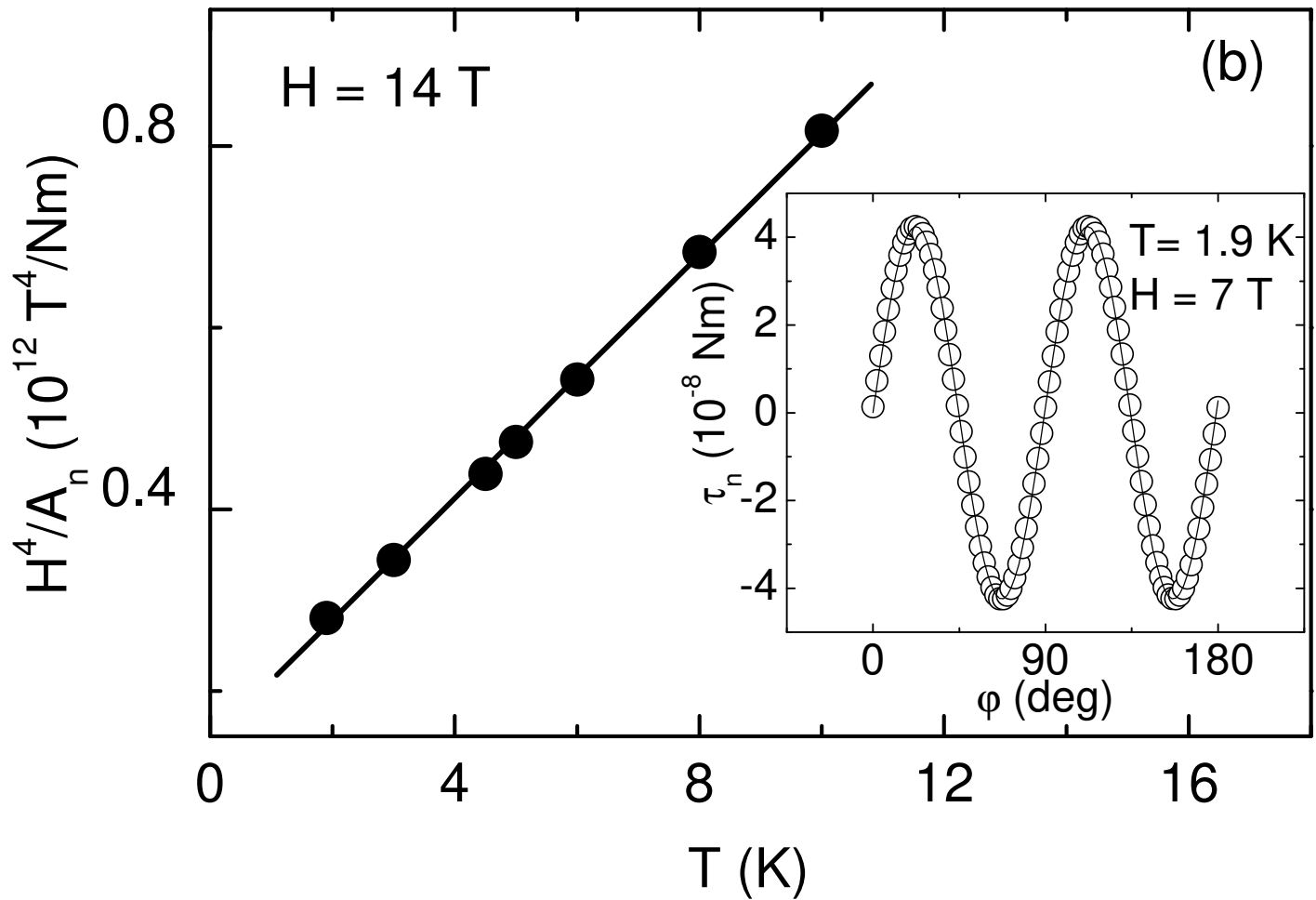
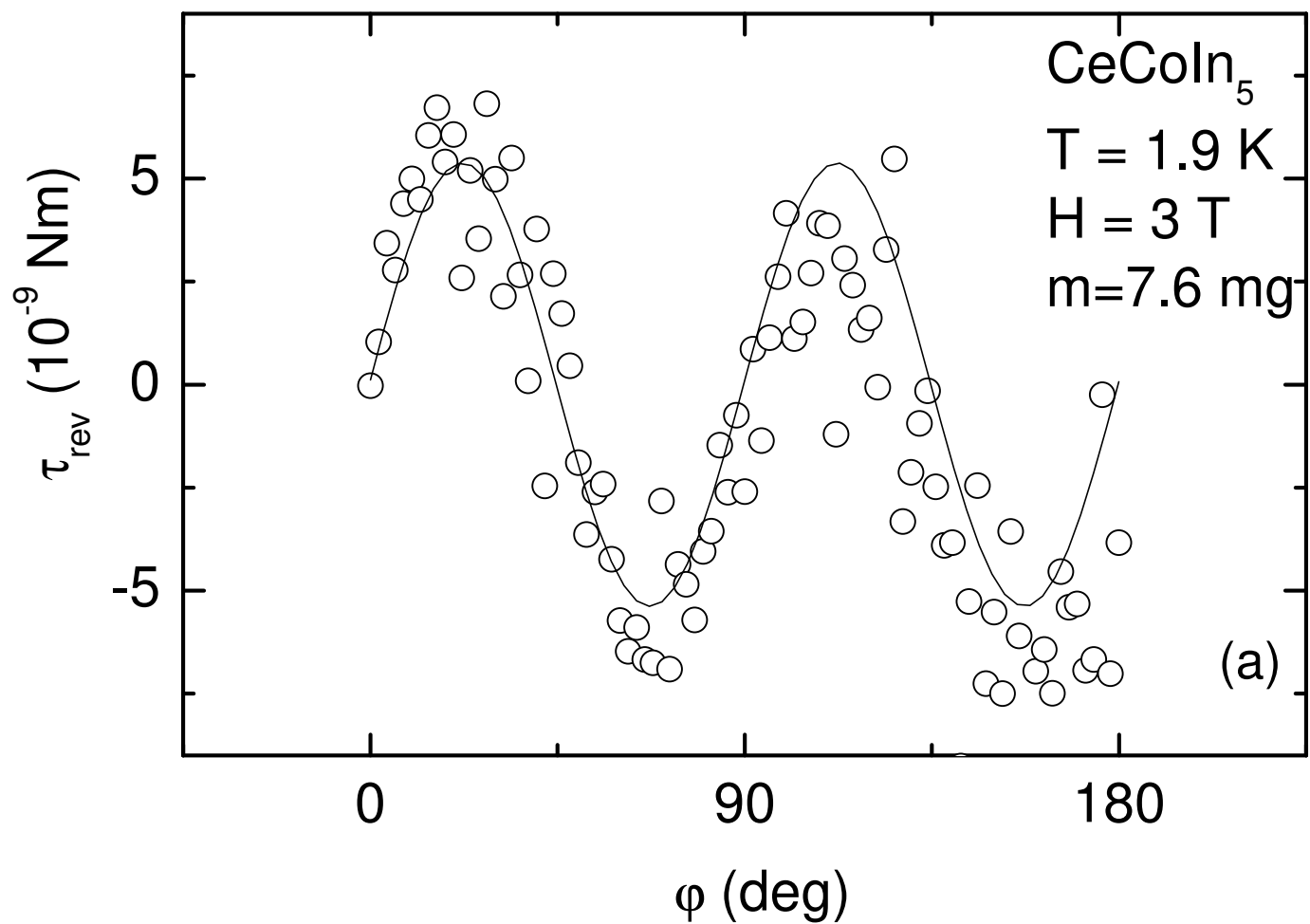
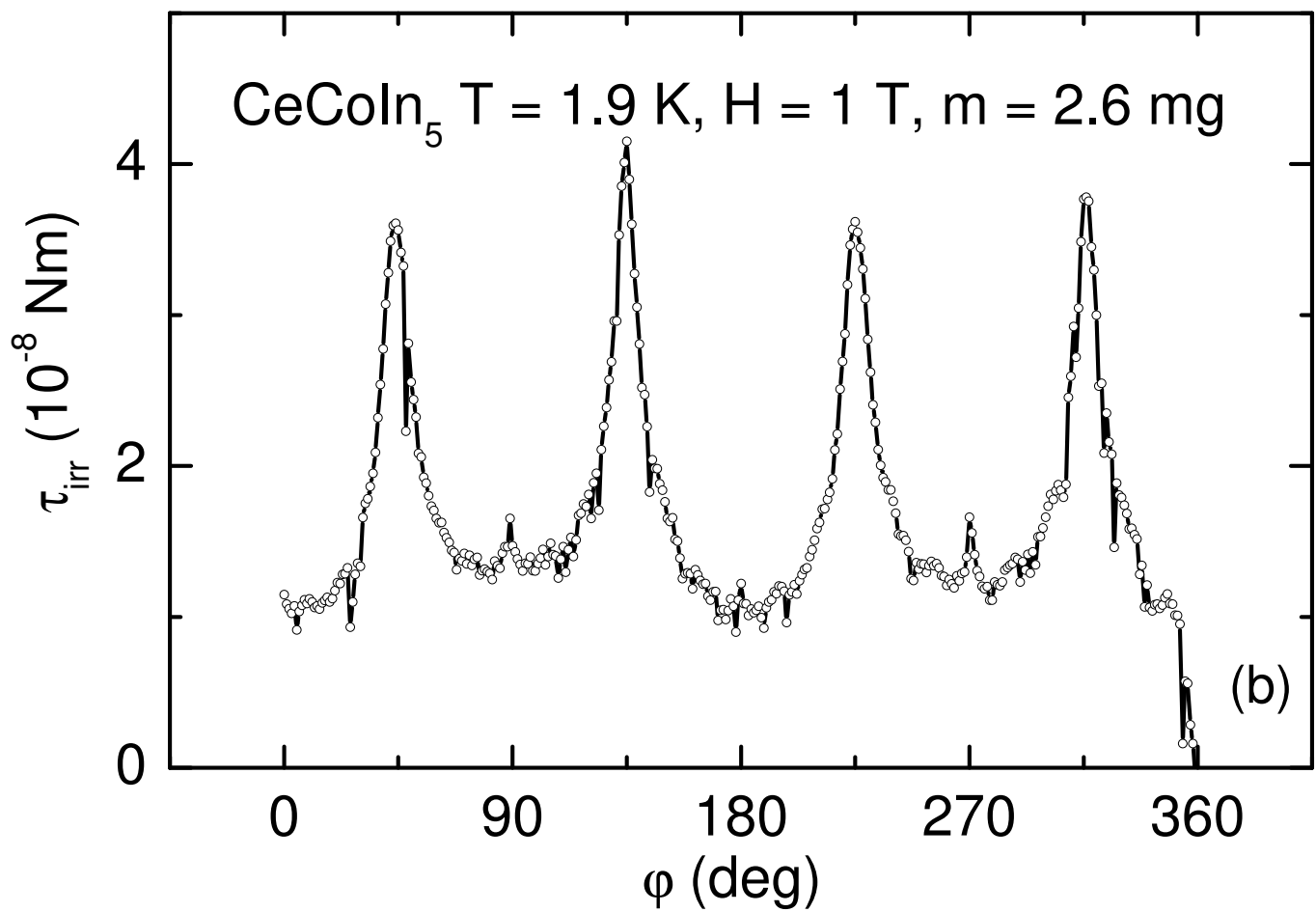
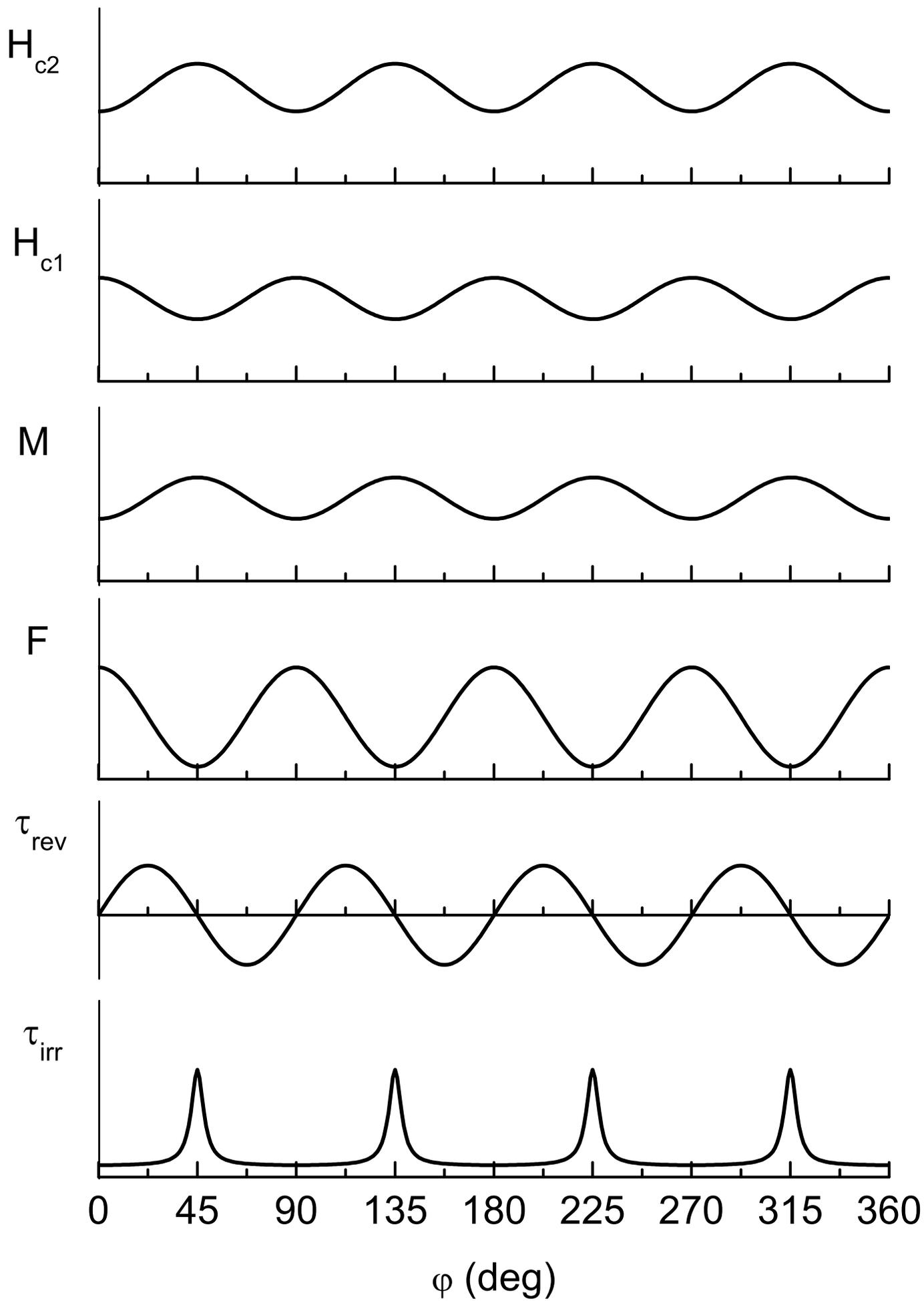
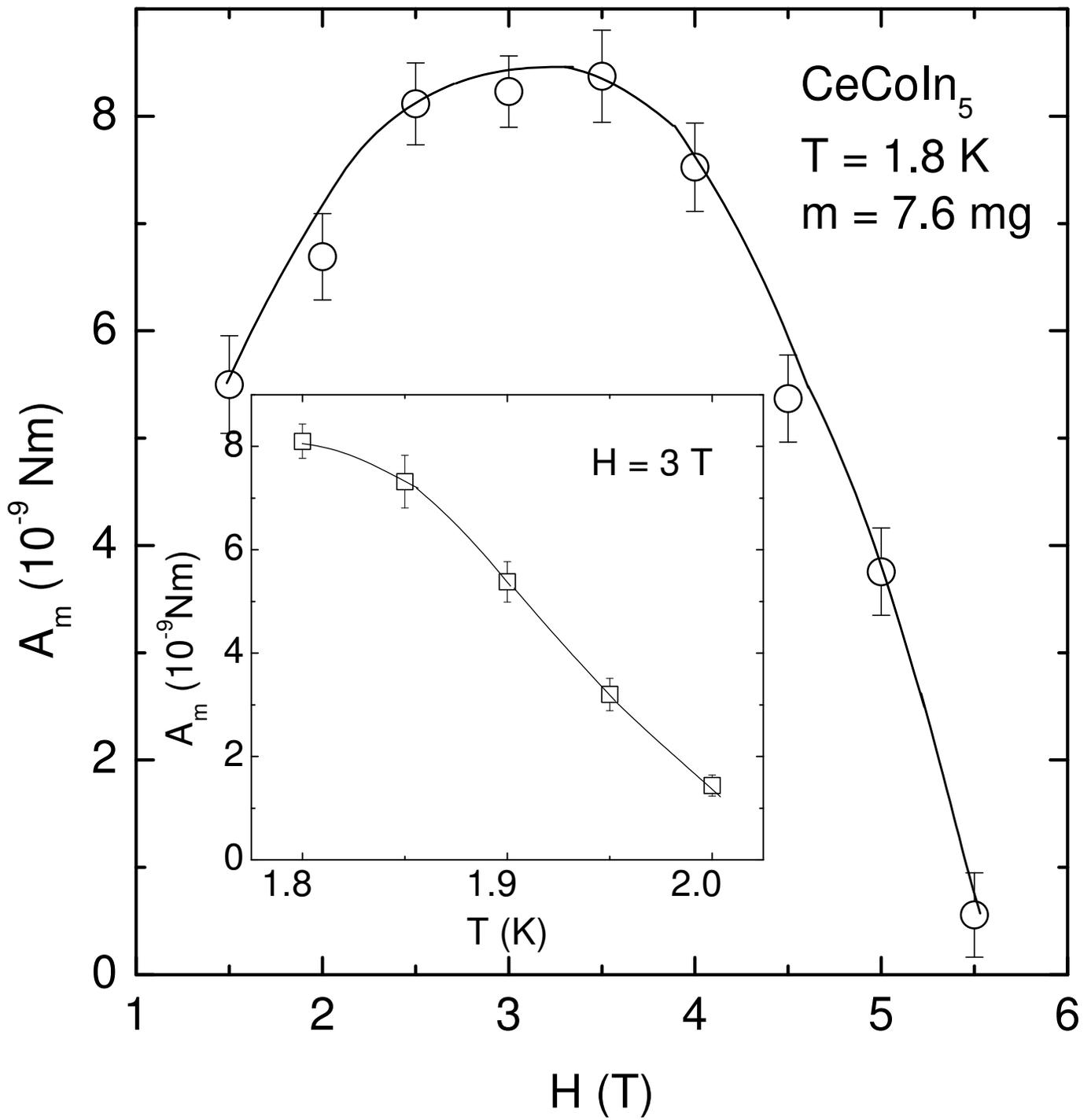
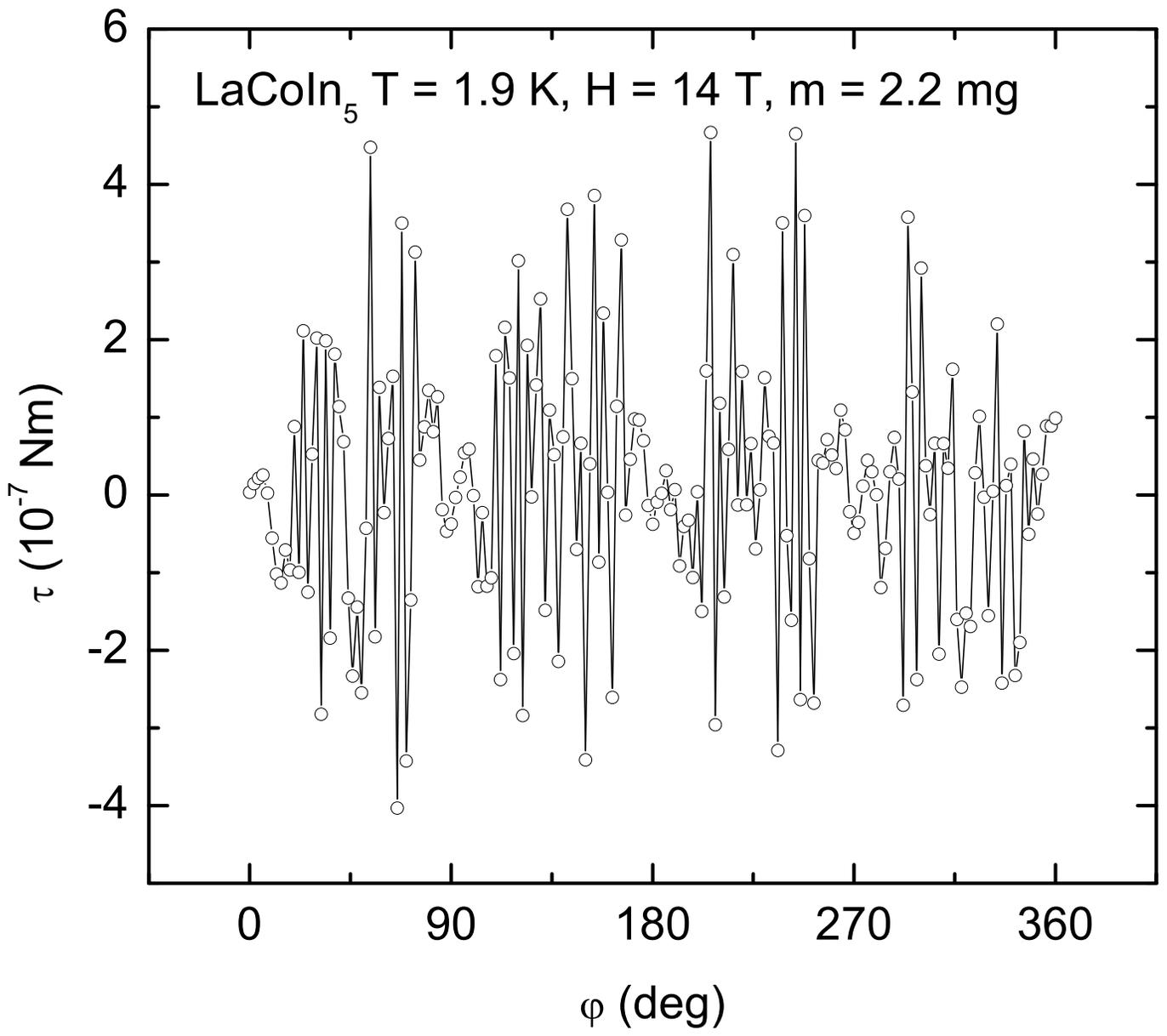
\end{document}